\documentclass[12 pt]{article}
\usepackage{latexsym}
\begin{document}
\begin{flushright} ULB--TH--00/03, UMH--MG--00/01\\
hep-th/0002156\\ February 2000\\
\end{flushright}
\begin{center} {\large THE FEFFERMAN-GRAHAM AMBIGUITY\\ \vspace{.1cm} AND
AdS BLACK HOLES}\\
\vspace{.5cm} K.~Bautier${}^{a,}$\footnote{ E-mail : kbautier@ulb.ac.be},
F.~Englert${}^{b,}$\footnote{ E-mail : fenglert@ulb.ac.be},
M.~Rooman${}^{b,}$\footnote{
E-mail : mrooman@ulb.ac.be} and Ph.~Spindel${}^{c,}$\footnote{ E-mail :
spindel@umh.ac.be}\\
\vspace{.3cm} {${}^a${\it Service de Physique Th\'eorique et Math\'ematique}}\\
{\it Universit\'e
Libre de Bruxelles, Campus Plaine, C.P.231}\\ {\it Boulevard du Triomphe,
B-1050 Bruxelles, Belgium}\\ 
\vspace{.2cm} {${}^b${\it Service de Physique Th\'eorique}}\\ 
{\it Universit\'e
Libre de Bruxelles, Campus Plaine, C.P.225}\\ {\it Boulevard du Triomphe,
B-1050 Bruxelles, Belgium}\\ 
\vspace{.2cm} {${}^c${\it
M\'ecanique et Gravitation}}\\ {\it Universit\'e de Mons-Hainaut, 20
Place du Parc}\\ 
{\it 7000 Mons, Belgium}\\ \end{center}

\begin{abstract} Asymptotically anti-de Sitter space-times in pure gravity
with negative cosmological constant are described, in all  space-time
dimensions greater than two, by classical degrees of freedom on the conformal
boundary at space-like infinity. Their effective boundary action has a
conformal anomaly for even dimensions and is conformally invariant for  odd
ones.   These degrees of freedom are encoded in traceless tensor fields in
the Fefferman-Graham asymptotic metric for any choice of conformally flat
boundary and generate all Schwarzschild and Kerr black holes in anti-de
Sitter space-time. We argue that these fields describe components of an
energy-momentum tensor of a   boundary theory and show
explicitly how this is realized in 2+1 dimensions. There, the
Fefferman-Graham fields reduce to the generators of the Virasoro
algebra and give the mass and the angular
momentum of the BTZ black holes. Their local expression is the Liouville
field in a general curved background.
\end{abstract}
\addtocounter{footnote}{-4}
\newpage

The conjectured equivalence, in the string theory approach to quantum
gravity, between supergravity in $D$-dimensional anti-de Sitter
space-time $AdS_D$ and some quantum conformal field theory living on its
boundary has drawn considerable interest \cite{maldacena,gubser,witten}.
It has been viewed as a manifestation of the holographic principle for
quantum degrees of freedom \cite{thooft,susskind,susswitt}. In this paper, we
show that there are {\em classical } degrees of freedom on the
conformal  boundary of $AdS$ and we discuss their relation with the
Fefferman-Graham description of gravity with negative cosmological constant.
These degrees of freedom   generate all $AdS$ Schwarzschild and Kerr black
holes. Their classical effective boundary action is conformally invariant
for odd dimensional space-time boundaries and presents the well known
conformal anomaly of gravity for even ones \cite{skenderis}. We derive here
this anomaly for general asymptotic anti-de Sitter space-times  from local
transformations of the boundary, avoiding ambiguities arising from global
transformations. These considerations are applied to the particular
case of $AdS_3$ where these degrees of freedom are described locally by a
Liouville field theory \cite{vandriel} on an arbitrary curved boundary
\cite{paris,weimar,skenderis2}.

In order to motivate our definition of asymptotic anti-de Sitter spaces,
we first review the anti-de Sitter geometry. The anti-de Sitter space-time
$AdS_D$ in $D=d+1$ dimensions is the hyperboloid of radius $l$
\begin{equation} X^2_1+.....+X^2_{d-1}+X^2_d-X^2_0-X^2_{-1} = -l^2
\label{hyperboloid} \end{equation} embedded in the $d+2$ flat space with
metric \begin{eqnarray} ds^2= dX^2_1+.....+dX^2_{d-1} -dX^2_0 -dUdV \, ,
\label{metric}\\ X_{-1}+X_d= U \, ,\nonumber\\ X_{-1}-X_d = V\, .
\nonumber \end{eqnarray} Its isometry group is $O\, (2,d)$. In terms of
the coordinates
$x_i= X_i/U$ and $U, U\neq 0$, the induced metric on the hyperboloid is
\begin{equation} ds^2 = U^2 ( dx^2_1+.....+dx^2_{d-1}-dx^2_0) + l^2
{dU^2\over U^2} \, . \label{bulk}
\end{equation} One may take as boundary of the hyperboloid, the
asymptotic projective cone described by Eq.(\ref{hyperboloid}) with
$l=0$. It can be parameterized, for $U \neq0$, by
\begin{eqnarray} ds^2_\infty = U^2 ds^2_B\, ,\label{cone} \\ds^2_B = (
dx^2_1+.....+dx^2_{d-1}-dx^2_0)\, ,
\label{infinity}
\end{eqnarray} and we shall refer to the geometry defined by $ds^2_B$ as a
boundary geometry. It is straightforward to check that the isometries $O\,
(2,d)$ leaving the metric Eq.(\ref{cone}) invariant are realized in the
boundary Minkowskian space-time Eq.(\ref{infinity}) (compactified by the
addition of the point at infinity corresponding to
$U=0$) by the conformal group in $d$ dimensions. Rotations $O\, (1,d-1)$
leave $U$ invariant while the dilations, the $2d$ translations and
special conformal transformations require a compensating transformation
on $U$. For $d=2$ the invariance of $ds^2_\infty$ under conformal
transformations of the boundary space-time is larger than $O\, (2,2)$ and
is generated by two copies of the Virasoro algebra.

Spaces which, at spacelike infinity, can be  described locally by the metric
Eq.(\ref{cone}) will be called asymptotic
$AdS$ space-times.
This definition has however to be made precise by specifying the limiting
procedure. To this effect, it is useful to enlarge the set of boundary
geometries $ds^2_B$ to the class of conformally flat spaces. For all
$d\geq 2$, the metric Eq.(\ref{cone}) is then invariant for the extended
conformal group $\cal C$  which leaves invariant
the conformal space of boundary metrics;  $\cal C$  contains as subgroups the
d-dimensional reparametrization group and  the Weyl group \cite{frw}.
The group $\cal C$ can be generated by bulk diffeomorphisms in the following
way. By writing $U^2= y^{-1}$, the metric Eq.(\ref{bulk}) takes the
Fefferman-Graham form
\cite{fefferman} \begin{equation} ds^2 = g_{yy} dy^2 +g^{(d)}_{\, ij}
dx^i dx^j = {l^2 dy^2\over 4y^2} + {1\over y} \tilde g_{ij}(x,y) dx^i
dx^j\,
\label{feff}
\end{equation} where the $d$-dimensional metric $\tilde g_{ij}(x,y)$ is
simply the $y$-independent Minkow\-skian boundary metric
Eq.(\ref{infinity}). We now take $\tilde g_{ij}(x,y)$ to be any metric
which tends, when $y\to 0$, to a conformally flat metric $\tilde
g_{(0)ij}( x)$. The diffeomorphisms keeping the form Eq.(\ref{feff}) are
those for which the Lie derivative of $g_{yy}$ and $g_{yi}$ vanish. They
are given by \cite{yankielowicz}:
\begin{eqnarray}
\delta y &=& 2\sigma(x) y\, ,
\label {diff}\\
\delta x^i &=& -\frac{l^2}{2}\int^y_0 \tilde g^{ij}(x,y^{\, \prime})
\sigma(x)_{,j} dy^{\, \prime} + \phi^i(x)\, ,\quad \phi^i(x)\, \hbox
{arbitrary}\, ,
\label{diff2}
\end{eqnarray} and induce on $\tilde g_{(0)ij}( x)$, up to a
reparametrization engendered by $\phi_i(x)$, the Lie variation:
\begin{equation}
\delta \tilde g_{(0)ij}( x) = -2\sigma(x) \tilde g_{(0)ij}( x) \, .
\label{conf}
\end{equation} Thus these diffeomorphisms form the group $\cal C$ acting
on the $AdS$ boundary metrics. Compensating the transformation
Eq.(\ref{conf}) by a Weyl transformation, we recover the invariance of
the metric Eq.(\ref{cone}) for the extended conformal group $\cal C$.
Note that this result follows from geometry and does not require any
gravitational field equations.

Given the conformally flat metric $\tilde g_{(0)ij}(x)$, it is possible
to reconstruct the bulk metrics of the type Eq.(\ref{feff}) which satisfy
in a neighborhood ${\cal V}$ of $y=0$  Einstein equations for pure gravity
with negative
cosmological constant $\Lambda \equiv -d(d-1)/l^2$. The reconstruction is
however not unique and yields, in addition to $AdS$, distinct asymptotic
anti-de Sitter space-times.

In the metric Eq.(\ref{feff}), the Einstein equations
\cite{skenderis,muck} take the form
\begin{equation} l^2 \tilde R^i_{\, j} + (d-2) h^i_{\, j}+ h^k_{\,
k}\delta^i_{\, j} - y(2\partial_y h^i_{\, j}+h^k_{\, k}h^i_{\, j}) =0
\label{einstein}
\end{equation} where
\begin{equation} h^i_{\, j} = \tilde g^{ik} \partial_y\tilde g_{kj}\, ,
\end{equation} together with the $y$-lapse constraint \begin{eqnarray}
R^y_{\ y} - {1\over2} R - d(d-1)/2l^2 &=&0 \nonumber\\ or \qquad l^2
\tilde R + 2(d-1) h^i_{\, i} + y(h^i_{\, j}h^j_{\, i}-h^i_{\, i}h^j_{\,
j}) &=&0\, ,
\label{constraint1}
\end {eqnarray} and the $d$ $y$-shift constraints
\begin{eqnarray} R^y_{\ j}&=&0\
\nonumber\\ or \qquad h^{i}_{\, j;i} - h^{k} _{\, k,j}&=&0 \, .
\label{constraint2}
\end {eqnarray}

Eq.(\ref{einstein}) can be solved iteratively by expanding $\tilde
g_{ij}(x,y)$ in a power series expansion in $y$. The generic term in the
expansion is $\tilde g_{(2k)}(x)\, y^k$ up to $k= d/2 -1$ for $d$ even
and to $(d-1)/2$ for $d$ odd. The coefficients of this expansion are
obtained algebraically in terms of $\tilde g_{(0)}(x)$ up to $\tilde
g_{(d-2)}(x)$ for
$d$ even and $\tilde g_{(d-1)}(x)$ for $d$ odd. For $d$ even, there may
be, in addition of $\tilde g_{(d)}(x)\, y^{d/2}$ a term $ \tilde
k_{(d)}(x)\, y^{d/2}\ln y$ where $\tilde k_{(d)}$ is traceless, namely
$Tr\, \tilde g_{(0)}^{-1} \tilde k_{(d)}=0$. While $\tilde k_{(d)}(x)$ is
still determined by $\tilde g_{(0)}(x)$, the traceless part $\tilde
g_{(d)}^t(x)$ of $\tilde g_{(d)}(x)$ is algebraically undetermined. For
$d$ odd, there is also a term $\tilde g_{(d)}^t(x)\, y^{d/2}$, where
$\tilde g_{(d)}^t(x)$ is traceless and is algebraically undetermined
\cite{fefferman,graham}. We shall refer to these undeterminacies as
the Fefferman-Graham ambiguity. All higher order terms in the expansion
can be expressed algebraically in terms of $\tilde g_{(0)}(x)$ and
$\tilde g_{(d)}^t(x)$.
To determine a solution in
$\cal V$, one must specify the transverse traceless fields $\tilde
g_{(d)}^t(x)$: although algebraically undetermined, these fields must
satisfy the differential equations obtained by  expanding in $y$ the
$y$-shift constraint Eq.(\ref{constraint2}).  Note that
Eq.(\ref{constraint1}) does not introduce new information.

We define as asymptotic anti-de Sitter space-time $\cal G$ any space-time
which can in 
${\cal V}$ be parameterized by a Fefferman-Graham metric
Eq.(\ref{feff}) such that: \\
$(i)$ $\tilde g_{(0)ij}( x) = \lim_{y\to 0} \tilde
g_{ij}(x,y)$ is conformally flat, \\ 
$(ii)$ the following boundary fields:
\begin{eqnarray} &\tilde g_{(2)}(x),
\tilde g_{(4)}(x)&.......\tilde g_{(d-1)}(x) \quad d\ odd\, ,
\label{boundodd}\\ &\tilde g_{(2)}(x), \tilde g_{(4)}(x)&.......\tilde
g_{(d-2)}(x), Tr\,\tilde g_{(d)}(x), \tilde k_{(d)}(x) \quad d\ even\, ,
\label{boundeven}
\end{eqnarray}
are expressed algebraically in terms of $\tilde g_{(0)}(x)$ in accordance
with Eq.(\ref{einstein})\footnote{The work of reference
\cite{yankielowicz} suggests that
these boundary conditions are in fact a direct consequence of the
transformations Eqs.(\ref{diff}) and (\ref{diff2}).}. 
While Eqs.(\ref{boundodd}) and (\ref{boundeven}) constitute boundary
conditions,
the algebraically undetermined fields $\tilde g_{(d)}^t(x)$ 
encode boundary dynamical degrees
of freedom whose nature will be  examined later. We first show that there
exists a finite action functional for these degrees of freedom and derive
its properties under the conformal group $\cal C$.

The Einstein-Hilbert action with suitable boundary terms is, in $\cal V$,
\begin{eqnarray} S&=&S_0 +{1\over 16 \pi G}\int_{\cal V} \sqrt {\vert g
\vert} (R+{d(d-1)\over l^2}) d^{d+1}x+ {1\over 8 \pi G} I \ ,
\label{action}\\ I&\equiv&-\int_{\cal S}\sqrt {\vert g^{(d)} \vert} (
K-C) d^dx \ .
\label{surface}
\end{eqnarray} where ${\cal S}$ is a $y=\bar y $ , $d$-dimensional,
boundary in $\cal V$ with topology $S_{d-1}\times R$ and $S_0$ is the
contribution to $S$ outside $\cal V$. The surface term $I$ is introduced,
as usual, to render the action stationary for solutions of Einstein's
equations with fixed fields on the boundary ${\cal S}$ \cite{gibbons}.
$K=K^i_{\, i}$ where $ K_{ij}$ is the extrinsic curvature tensor in the
metric $g^{(d)}_{\, ij}$ of Eq.(\ref{feff}). The constant $C$ is
introduced for convenience and will be fixed later.

We perform a diffeomorphism in
$D$ dimensions defined by an infinitesimal displacement field $\xi^\mu
(x,y)$ which vanishes outside $\cal V$ . To compute the corresponding
variation of the action Eq.(\ref{action}) we first write the surface
integral in covariant form by introducing the normal D-vectors
$n_\mu$ to the surface elements $d\Sigma_\mu $ on $ {\cal S}$. By
embedding the displaced surface ${\cal S}$ under the diffeomorphism in a
family of surfaces characterized by infinitesimal displacements
$\alpha\, \xi^\mu$ where $\alpha$ varies from $0$ to $1$, one defines a
field $n_\mu$ normal to the surfaces
$\alpha=$ constant. The extrinsic curvature scalar $K$ can be written as
a D-dimensional scalar $K\equiv  - n^\mu_{\ ;\mu}$. Using \begin{equation}
\sqrt{\vert g^{(d)}\vert} d^dx = \sqrt {\vert g \vert} n^\mu d\Sigma_\mu
\, , \label{element}
\end{equation} the integral Eq.(\ref{surface}) becomes \begin{equation}
I=\int_{ \cal S} \sqrt {\vert g \vert} n^\mu (n^\sigma_{\;\sigma}+C)
d\Sigma_\mu\ .
\label{surfacecovariant}
\end{equation} The variation of the action is obtained by taking the Lie
derivative of the integrand of volume terms after transforming the
surface term to a volume integral of a D-divergence. We get
\begin{equation}
\delta_\xi S = {1\over 16 \pi G}\int_{\cal S}\sqrt {\vert g
\vert} \{(R+{d(d-1)\over l^2}) + 2 [n^\mu (n^\sigma_{\
;\sigma}+C)]_{;\mu}\}\xi^\rho d\Sigma_\rho\, . \ \label{lie}
\end{equation}
We substitute in Eq.(\ref{lie}) the identity \begin{equation} (n^\mu
n^\sigma_{\ ;\sigma})_{;\mu} = n^\mu_{\ ;\mu} n^\sigma_{\ ;\sigma} -
n^\mu_{\ ;\sigma}n^\sigma_{\ ;\mu} - n^\mu n^\sigma R_{\sigma\mu} +
(n^\mu n^\sigma_{\ ;\mu})_{;\sigma}\, . \label{normal} \end{equation} It
is easily checked that on ${\cal S}$ the last term in Eq.(\ref{normal})
gives no contribution. If we insert  the $y$-lapse
constraint Eq.(\ref{constraint1}) in the Gauss-Codazzi equation
\begin{equation} 2G^y_{\ y} \equiv 2R^y_{\ y} - R = - R^{(d)} + [K^i_{\, i}
K^j_{\, j} - K^i_{\, j}K^j_{\, i}]
\end{equation} and use the identity
$\sqrt {\vert g
\vert}\xi^\rho d\Sigma_\rho =\sqrt{\vert g^{(d)}\vert}\xi d^dx$ with $\xi
=\xi^\mu n_\mu$, we get
\begin{equation}
\delta_\xi S = {1\over 8 \pi G}\int_{\cal S}\sqrt {\vert
g^{(d)} \vert} \{ R^{(d)} + {d(d-1)\over l^2} + 2C K \}\xi d^dx\, .
\label{varied} \end{equation} Taking $C= (d-1)/l$, this can be written as
\begin{equation} \delta_\xi S = {l\over 16 \pi G}\int_{\cal
S}\sqrt {\vert \tilde g^{(d)}
\vert} ( R^{(d)} + {d-1\over l^2 } h^i_{\, i}\, ) /\bar y^{d/2 }\,\delta
\bar y \,d^dx\, . \label{liefinal} \end{equation} Eq.(\ref{liefinal}) is
similar to the equation obtained in reference \cite{muck} for the
variation of the Einstein-Hilbert action but there are two noticeable
differences. First, our result is valid  for arbitrary {\em local} variations
$\delta \bar y$ around the surface $y=\bar y$. Second, as shown below, in
the limit ${\bar y}\to 0$ Eq.(\ref{liefinal}) does not require the
evaluation of
the action Eq.(\ref{action})  on a solution of Einstein's
equations but only on an arbitrary asymptotically anti-de Sitter space-time
$\cal G$.

We  examine Eq.(\ref{liefinal}) when $\bar y$ tends to zero. The
integral is divergent and we may classify the divergences by expanding
the integrand in power series in $\bar y$. We expand the numerator of the
integrand of Eq.(\ref{liefinal}), keeping only non vanishing terms as
$\bar y$ goes to $ 0$. These divergences arise from terms of order less
than
$\bar y^{d/2 }$ in the expansion of $\sqrt {\vert \tilde g^{(d)} \vert}$,
$ R^{(d)}$ and $h^i_{\, i}$, and potentially from terms in $h^i_{\, i}$
(which contains a $\bar y$-derivative) of order $\bar
y^{d/2}$ or $ \bar y^{d/2}\ln \bar y$ in the expansion of the metric
$\tilde g_{ij}(x,y)$. However, due to tracelessness, there is no
contribution containing $\tilde g_{(d)}^t$. (For the same reasons, there
is no contribution from the logarithmic term for $d$ even.) The
divergent contributions depend only on the fields appearing in
Eqs.(\ref{boundodd}) and (\ref{boundeven}) for which the $y$-constraint
used
in the derivation of Eq.(\ref{liefinal}) reduces to an identity.  Thus,
as announced, in the limit ${\bar y}\to 0$, this equation is valid for all
asymptotic anti-de Sitter space-times $\cal G$. The divergent terms
have no dynamical content.

We denote by $S(\cal G)$ the action $S$ evaluated on a space $\cal G$.
Integrating Eq.(\ref{liefinal}) for a global variation $\delta \bar y$ one
gets when
$\bar y \to 0$
\begin{eqnarray} {S(\cal G)}&=& {A_{(d/2)-1}\over \bar y^{(d/2)-1}} +...+
{A_{1/2}\over \bar y^{1/2}} + {S(\cal G)}_{fin}\, ,\qquad d\ odd
\label{odd}\\ {S(\cal G)}&=& {A_{(d/2)-1}\over \bar y^{(d/2)-1}} + ..+ A_0
\ln \bar y + {S(\cal G)}_{fin}\, ,
\qquad d\ even \label{even}
\end{eqnarray} where $ {S(\cal G)}_{fin}$ is independent of $\bar y$ and
finite. The fields $\tilde g_{(d)}^t(x)$ enter only $ S_{fin}({\cal G})$ which
contains therefore all the dynamics. One may view $ S_{fin}({\cal G})$ as
an effective boundary action for these dynamical degrees of freedom.

 Comparing Eqs.(\ref{odd}) and (\ref{even}) with Eq.(\ref{liefinal})
for a local variation of $\delta \bar y$, we see that the coefficients of
the divergent terms $A$ can be expressed as  surface integral of local
functions of $\tilde g_{(0)}(x)$ only. In particular we may write
\begin{equation} A_0=\frac{1}{2}
\int \sqrt {\vert g^{(d)} \vert} {\cal A}(\tilde g_{(0)}(x)) d^dx,
\end{equation} where $\sqrt {\vert g^{(d)} \vert} {\cal A}/2$ is  the
coefficient of $\delta \bar y /\bar y$ in
Eq.(\ref{liefinal}).

Taking for $\delta \bar y$ the particular local transformation
Eq.(\ref{diff}) renders  the contribution of the $\cal A$ term in
Eq.(\ref{liefinal}) finite. The  unsubtracted action
$S({\cal G})$ is invariant under the combined  Lie transformation Eq.(\ref
{liefinal}) and (minus) the conformal  transformation
of  Eq.(\ref{conf}) acting on $\tilde g_{(0)}(x)$, provided the conformal
transformation of the field
$\tilde g_{(d)}^t(x)$ is defined by (minus) the Lie variation under the
diffeomorphism Eqs.(\ref{diff}) and (\ref{diff2}).  This expresses the
invariance of $ds^2_\infty$ under the group $\cal C$. The local divergent
integrals are separately invariant under these transformations except for the
logarithmic divergent integral $A_0$ which is conformally invariant. The
compensation
of the finite $\delta \bar y / \bar y$ term identifies ${\cal
A}(\tilde g_{(0)}(x))$ to the well known anomaly  of the ``renormalized''
action $ S_{fin}({\cal G})$  for $d$ even while for $d$ odd $ S_{fin}({\cal
G})$ is conformally invariant \cite{witten,skenderis}. As our
proof uses {\em local} variations of $\bar y$, it is free of ambiguities
hidden in proofs based on global variations and its validity extends to all
  asymptotic anti-de Sitter space-times $\cal G$.

We now discuss the nature of the
boundary degrees of freedom encoded in
$\tilde g_{(d)}^t(x)$.

When $\cal G$ solves the Einstein's equations,
$\tilde g_{(0)}(x)$ can still be taken as an arbitrary conformally flat
metric but
$\tilde g_{(d)}^t(x)$ must satisfy all Einstein's equations, which means
that it must solve the $y$-shift constraints Eq.(\ref{constraint2}).
Expanding
$h^{i}_{\, j;i} - h^{k} _{\, k,j}$ in a power series in $y$, all terms of
order less than
$y^{d/2 -1}$  vanish identically because they contain only coefficients
$\tilde g_{(2k)}(x)$ which are already determined algebraically in terms of
$\tilde g_{(0)}(x)$. The term in $y^{d/2 -1}$ yields a differential equation
for
$\tilde g_{(d)}^t(x)$. From Eq.(\ref{constraint2}), we can see that this
equation takes the form
\begin{equation} D_i\, \tilde g_{(d)j}^{\ i} + \psi_j[\tilde g_{(0)}(x)]
=0 \label{freedom}
\end{equation} where here and in what follows, indices are raised with
the boundary metric $\tilde g_{(0)}(x)$ and
$D_i$ is the covariant derivative in the same metric. The
quantities $\psi_i[\tilde
g_{(0)}(x)]$ contain only terms determined in terms of derivatives of
$\tilde g_{(0)}(x)$, hence expressible in terms of curvature terms of the
boundary metric, and depend explicitly on $d$. Note that
Eq.(\ref{freedom}), which gives equations of motion for the degrees of
freedom on the boundary of $AdS_{d+1}$, reduces to an identity for all
higher dimensional boundaries, when the explicit dependence of
$\psi_i[\tilde g_{(0)}(x)]$ on the dimension is taken into account.

We illustrate Eq.(\ref{freedom}) for the cases $d=2,3$ and $4$. When
$d=2$ the order zero in $y$ in Eq.(\ref{constraint2}) yields
\begin{equation} D_i(\tilde g_{(2)j}^{\ i} + \delta^i_{\
j}\frac{l^2}{2}\tilde R) = 0\, , \label{constraint2d2}
\end{equation} where we have used Eq.(\ref{einstein}) to get the trace
part of $\tilde g_{(2)}$ in terms of $\tilde g_{(0)}(x)$:
\begin{equation} \tilde g_{(2)i}^{\ i}= -\frac{l^2}{2}\tilde R\, .
\label{einsteind2} \end{equation} For $d=3$ we take the derivative of
Eq.(\ref{constraint2}) with respect to $y^{1/2}$; the Christoffel symbols
are not affected at this order and taking into account the vanishing of
the trace $ g^i_{(3)i}$, we simply get
\begin{equation} D_i\, \tilde g^{\ i}_{(3) j} =0\, .
\label{freedom3}
\end{equation} For $d=4$, we take the derivative with respect to $y$ and
take into account the change of the Christoffel symbols in the covariant
derivative. We get \begin{equation} D_i\, \{\tilde g^{\ i}_{(4) j} +
\frac{l^4}{16} [\tilde R \tilde R^i_{\, j} -2 \tilde R^{im}\tilde R_{mj}
+ \frac{1}{2}\delta^i_{\, j} (\tilde R^{mn}\tilde R_{mn}
-\frac{5}{9}\tilde R^2)]\}=0\, . \label{freedom4}
\end{equation} The form of Eq.(\ref{constraint2d2}), (\ref{freedom3}) and
(\ref{freedom4}) can be summarized by
\begin{equation} D_i (\tilde g^{\ i}_{(d)j} + \xi^i_{(d)j})=0
\label{freedomd}
\end{equation} where $\xi^{\ i}_{(d)j}$ is constructed out of the
curvature tensors and their derivatives. From the conservation law
Eq.(\ref{freedomd}) we may define a conserved tensor
\begin{equation} T^i_{(d)j}=\frac{1}{\alpha_d} (\tilde g^{\ i}_{(d)j} +
\xi^{\ i}_{(d)j}) \label{trace}
\end{equation} where
$\alpha_d$ is a numerical coefficient. Choosing $\alpha_2 =8\pi Gl$ and
$\alpha_4=4\pi Gl$, we get
\begin{eqnarray} T^i_{(2)i}&=& \frac{3l}{2G}\frac{\tilde R}{24\pi},
\label{trace2}\\ T^i_{(3)i}&=&0, \label{trace3}\\
T^i_{(4)i}&=&\frac{l^3}{8\pi G}(\frac{1}{8} \tilde R^{mn}\tilde R_{mn}
-\frac{1}{24}\tilde R^2). \label{trace4} \end{eqnarray} We see that the
traces of the covariantly conserved tensors $T^i_{(d)j}$, which are
related by Eq.(\ref{trace}) to the Fefferman-Graham fields
$\tilde g_{(d)}^t(x)$, reproduce for $d=2$ and $4$ the gravitational
anomalies
$\cal A$ computed from Eq.(\ref{liefinal}) and its vanishing for $d=3$.
Note that in flat space the equations of motion Eq.(\ref{freedom})
express already for all $d$ the conservation of the quantity
$\tilde g_{(d)j}^{\ i}$, because then $\psi_i[\tilde g_{(0)}(x)]$
vanishes. The extension of this conservation equation to the case of a
general boundary metric $\tilde g_{(0)}(x)$ relies on the precise
form of $\psi_i[\tilde g_{(0)}(x)]$ and does not involve the dynamical
fields $\tilde g_{(d)}^t(x)$.
These results suggests that, for all $d$, the degrees
of freedom hidden in the Fefferman-Graham ambiguity $\tilde g_{(d)}^t(x)$
can be expressed in terms of a conserved energy momentum tensor of some
boundary fields.  The trace  of this energy-momentum tensor
would, on the equations of motion, be equal to the trace anomaly
$\cal A$. These  boundary fields would describe  the shape of the surface
$\cal S$ in the limiting process ${\bar y} \to 0$.

The boundary degrees of freedom generate all
Schwarz\-schild and Kerr black holes in asymptotic $AdS_D$ space-time for
all $D$. The metric of $AdS$ Schwarzschild black holes of mass $M$ can be
taken to be
\begin{equation} ds^2 = - (1 - {\lambda\over r^{d-2}} +{r^2\over l^2})
dt^2 + (1 - {\lambda\over r^{d-2}}+{r^2\over l^2})^{-1} dr^2 + r^2
d\Omega_{d-1}^2\, , \label{schwar}
\end{equation} where $\lambda =\nu_D G_DM $. $G_D$ is the gravitational
constant in $D=d+1$ dimensions and
$\nu_D$ is a $D$-dependent numerical coefficient. When $r\to \infty$, the
relation between $r$ and the variable $y$ in the metric Eq.(\ref{feff})
is $dr/r\to - dy/2y$ or equivalently by suitable choice of the
integration constant constant, $r/l \to y^{-1/2}$ . Hence the leading
order in $y$ of the mass term in the coefficient of $dt^2$ is $O(d/2)$.
As the mass term has no counterpart in the pure $AdS_{d+1}$ geometry its
leading contribution to the expansion of $\tilde g_{ij}(x,y)$ is the
traceless quantity $\tilde g_{(d)}^t(x)$ which is here a constant
independent of $x$.

The tracelessness of the mass term can be verified explicitly. The
relation between $r$ and $y$ is defined by \begin{equation} (1 -
{\lambda\over r^{d-2}} +{r^2\over l^2})^{-1/2}dr= -{ldy\over 2y}\, .
\end{equation} Posing, for $d>2$, $2r/l = \xi- 1/\xi$, one obtains, up to
order $\xi^{-d}$, \begin{equation}
\ln \xi - {2^{d-1}\over l^{d-2} \lambda^2 d} \xi^{-d}= -{1\over2} \ln
{y\over 4}\, .
\end{equation} Expressing this equality to the same order in terms of
$r$, one gets \begin{equation} {r\over l} = {1\over y^{1/2}}(1+
{1\over4}y + {\lambda^2\over 2 l^{d-2} d} y^{d/2})\, , \end{equation} from
which one easily verifies that $ \tilde g^{ik}_{(0)}\tilde g_{(d), ki}$
vanishes. For $d=2$, the exact relation between $r$ and $y$ is obtained
along similar lines by the change of variable $2r/l = \xi+ 1/\xi$ and the
vanishing of $ \tilde g^{ik}_{(0)}\tilde g_{(2), ki}$ follows.

These conclusions can be extended to Kerr black holes. The dependence on
$r$ and hence of the leading order in $y$ of the mass term in
Eq.(\ref{schwar}) follows from dimensional considerations : $G_DM$ has
dimension $D-3=d-2$ and its contribution to $g_{tt}$ cannot depend on
$l$. Similarly the angular momentum $J$ will give a term proportional to
$G_DJ$ in $g_{t\phi}$ of the same dimensionality. This is the leading
dependence on the angular momentum in the Kerr metric and is again
encoded in the traceless field $\tilde g_{(d)}^t(x)$.

We now apply these considerations  to asymptotic $AdS_3$
space-times \cite{paris}. When $d=2$,  Eq.(\ref{even}) gives
 \begin{equation}
S({\cal G})=A_0 \ln \bar y + S_{fin}({\cal G})\, . \end{equation}
Eq.(\ref{trace2}) shows that the conformal anomaly, computed from
Eq.(\ref{liefinal}) (see also
\cite{tanii}) \begin{equation} {\cal A}= \frac{3l}{2G}\frac{\tilde
R}{24\pi} \label{central}
\end{equation} agrees with the trace of the
conserved energy momentum tensor $T^{\ i}_{(2) j}$ defined by
Eqs.(\ref{constraint2d2}) and (\ref{trace})
\begin{equation}
\tilde g_{(2)j}^{\ i} +\delta^i_{\ j}\frac{l^2}{2}\tilde R={8\pi Gl} 
T^{\ i}_{(2)j}\, .
\label{stress}
\end{equation} To verify that $T^{\ i}_{(2)j}$ is indeed the
energy-momentum tensor of a  boundary field, we consider the Weyl
transformation of $\tilde g_{(0)ij}(x)$ and $\tilde g_{(2)ij}(x)$ 
given by (minus) the Lie derivatives defined by the diffeomorphism 
Eqs.(\ref{diff}) and (\ref{diff2}). We have
\cite{yankielowicz} \begin{eqnarray}
\delta_{Weyl}\, \tilde g_{(0)ij}(x) &=& 
2\sigma(x) \tilde g_{(0)ij}(x)\, ,\\
\delta_{Weyl}\, \tilde g_{(2)ij}(x) &=& l^2(\sigma(x) )_{,i\, ;j}\, .
\label{weyl}
\end{eqnarray} Taking a scalar field $\phi$ such that $e^\phi$ has
conformal weight
$-1$ ($\delta_{Weyl}\, \phi=-\sigma(x) $), one obtains from
Eq.(\ref{weyl})
\begin{equation}
\tilde g_{(2)ij}=l^2[-\phi_{,i\, ;j}+\phi_{,i}\phi_{,j} +\tilde
g_{(0)ij}(\lambda e^{2\phi}-\frac{1}{2}\tilde g_{(0)}^{\,
kl}\phi_k\phi_l)]\, ,
\label{phi}
\end{equation} where we have included the general conformal invariant
term $\lambda g_{(0)ij} e^{2\phi}$, with arbitrary $\lambda$. This
equation identifies $T_{ij}$ in  Eq.(\ref{stress}) as the energy-momentum
tensor derived  from the Liouville action \begin{equation}
S_{Liouville}=\frac{-l}{8\pi  G}\int \sqrt{\vert \tilde g_{(0)} \vert}
(\frac{1}{2} \tilde g_{(0)}^{ij} \phi_{,i}\, \phi_{,j} +\frac{1}{2}\tilde
R \phi +\lambda e^{2\phi}) d^2x\, , \label{liouville}
\end{equation} computed on the Liouville equations of motion. As it can
be shown that Eq.(\ref{phi}) is integrable on the equations of motion,  we
have identified locally all the degrees of freedom
encoded in the Fefferman-Graham traceless field $\tilde g_{(2)}^t(x)$ to
the Liouville field on a arbitrary curved 2-dimensional background and the
Liouville action is locally equivalent to $S_{fin}({\cal G})$. The action
Eq.(\ref{liouville}) is the generalization to curved boundaries
\cite{weimar} of the realization given in reference
\cite{vandriel} of the Brown-Henneaux central charge \cite{brown} encoded in
Eq.(\ref{central}). The traceless components in Eq.(\ref{phi}) give the
mass and angular momentum
\cite{brown} of all the BTZ black holes \cite{btz1,btz2}
\begin{equation}
\frac{8\pi G}{l}T_{++}=\frac{1}{4}(M-J/l)\, ,\quad \frac{8\pi
G}{l}T_{--}=\frac{1}{4}(M+J/l)\, . \end{equation}

To conclude, we stress that, for asymptotic anti-de Sitter space-times,
the quantum holographic principle seems to have a classical ancestor.
What is left of this feature in the limit $\Lambda \to 0$ is unclear to
us.

\subsection*{Acknowledgements}

We thank Marc Henneaux for a very enlightening discussion. K.~B. is
``Chercheur F.R.I.A." and M.~R. is Senior Research Associate F.N.R.S.
This work has been partly supported by the ``Actions de
Recherche Concert{\'e}es" of the ``Direction de la Recherche
Scientifique - Communaut{\'e} Fran{\c c}aise de Belgique" and by
IISN - Belgium (convention 4.4505.86).

\end{document}